%% file: main.tex
\documentclass[conference]{IEEEtran}
\IEEEoverridecommandlockouts

\ifCLASSOPTIONcompsoc
  \usepackage[nocompress]{cite}
\else
  \usepackage{cite}
\fi

\usepackage[utf8]{inputenc}
\usepackage{chngcntr}
\usepackage[cmex10]{amsmath}
\usepackage{afterpage}
\usepackage[T1]{fontenc}
\usepackage{lettrine}
\usepackage{amsfonts}
\usepackage{amssymb}
\usepackage{graphicx}
\usepackage{csquotes}
\usepackage{mathtools}
\usepackage{bbm}
\usepackage[normalem]{ulem}
\usepackage{tabularx}
\usepackage[nottoc]{tocbibind}
\usepackage{float}
\usepackage[caption = false]{subfig}
\usepackage{enumerate}
\usepackage[dvipsnames]{xcolor}
\usepackage{tikz}
\usepackage{pgfplots}
\usepackage{qcircuit}
\usepackage{physics}
\tikzset{>=latex}
\usepackage{algorithm}
\usepackage[noend]{algpseudocode}
\algnewcommand\algorithmicforeach{\textbf{for each}}
\algdef{S}[FOR]{ForEach}[1]{\algorithmicforeach\ #1\ \algorithmicdo}

\usepackage{microtype}
\usepackage[shortcuts,acronym]{glossaries}
\usepackage{hyperref}
\usepackage[nameinlink]{cleveref}

\usetikzlibrary{shapes, arrows, positioning, external}
\pgfplotsset{compat=1.15}
\usepgfplotslibrary{groupplots}

\title{Quantum Autoencoders for Learning Quantum Channel Codes}
\date{}

\author{
\IEEEauthorblockN{ Lakshika Rathi$^1$, Stephen DiAdamo$^{1*}$, and Alireza Shabani$^2$}
\IEEEauthorblockA{
\textit{$^1$Cisco Quantum Lab, Garching bei M\"unchen, Germany}\\
\textit{$^2$Cisco Quantum Lab, Los Angeles, USA}
}
\thanks{$^{*}$Corresponding author: sdiadamo@cisco.com}
}

\begin{document}
\maketitle

\begin{abstract}

This work investigates the application of quantum machine learning techniques for classical and quantum communication across different qubit channel models. By employing parameterized quantum circuits and a flexible channel noise model, we develop a machine learning framework to generate quantum channel codes and evaluate their effectiveness. We explore classical, entanglement-assisted, and quantum communication scenarios within our framework. Applying it to various quantum channel models as proof of concept, we demonstrate strong performance in each case. Our results highlight the potential of quantum machine learning in advancing research on quantum communication systems, enabling a better understanding of capacity bounds under modulation constraints, various communication settings, and diverse channel models.

\end{abstract}

\section{Introduction}

In classical coding theory, machine learning (ML) has emerged as a powerful tool for generating communication codes that nearly achieve channel capacity for various channel models. Promising results using ML have been demonstrated for learning channel codes~\cite{o2017introduction, letizia2021capacity, qin2019deep, jiang2019turbo}. Specifically, in these works, autoencoders have been proposed as an alternative model for communication systems. An autoencoder is a type of neural network used for unsupervised learning. It consists of an encoder neural network that can compress the input data into a lower-dimensional representation, and a decoder neural network that reconstructs the original input from the compressed representation. The network learns to minimize the reconstruction error.

In these past works, the authors map the original models introduced by Claude Shannon~\cite{shannon1948mathematical} to layers of a neural network representing the encoder, the noisy channel, and the decoder. By framing the training problem as a classification problem, the neural network can learn how to overcome the channel noise while simultaneously compressing the data that is transmitted. Moreover, particularly in  \cite{letizia2021capacity}, the cost function is set such that the autoencoder is encouraged to increase the mutual information of the code to better train for a capacity achieving code.

In this work, our objective is to apply the insights gained from previous studies to the realm of quantum communication. Quantum autoencoder models have been swiftly adopted across various research domains \cite{tian2022recent, ngairangbam2022anomaly, romero2017quantum, huang2020realization, zhu2021quantum, bondarenko2020quantum}. In these contexts, quantum autoencoders draw inspiration from variational quantum algorithms~\cite{cerezo2021variational} which iteratively update parameters to learn how to compress and decompress the Hilbert spaces encompassing quantum systems. By operating within a compressed, latent space, these models achieve more efficient computations within smaller Hilbert spaces. This becomes particularly valuable during the NISQ era of quantum computing, where resource conservation is of high importance

\begin{figure}
    \centering
    \includegraphics[]{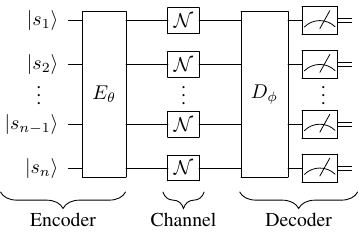}
    \caption{The circuit model for classical communication. A message $s\in\mathcal{M}$ is encoded into a series of qubits $\ket{s_i}$ and input to a parameterized encoder. Next, the channel effects $\mathcal{N}$ are applied. Once through the channel, a parameterized decoder is applied, and a series of outputs $\hat{s}_i, i \in [n]$ are collected.  }
    \label{fig:coder}
\end{figure}

\begin{figure}
    \centering
    \includegraphics[]{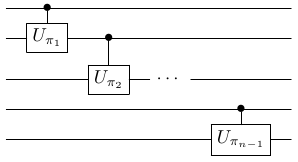}
    \caption{An example of a parameterized information-pooling circuit, where $U_{\pi_i}$ are arbitrary rotation gates parameterized by $\pi_i\in\mathbb{R}^3$. }
    \label{fig:pooler}
\end{figure}

\begin{figure}
    \centering
    \includegraphics[]{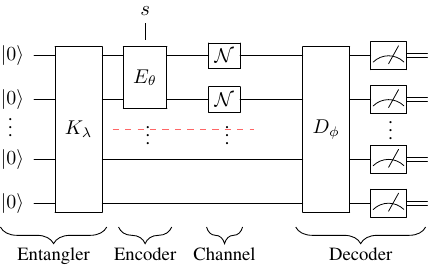}
    \caption{The circuit model for EA-classical communication. A parameterized circuit generates entanglement using learned entanglement resources. A message $s\in\mathcal{M}$ is fed into the encoder and half of the qubits are encoded by a parameterized encoder. Next, the channel effects are applied to half of the qubits. Once through the channel, a parameterized decoder is applied to the whole system, and a series of outputs $\hat{s}_i, i \in [n]$ are collected. The red dashed line represents the system separation.}
    \label{fig:ea-coder}
\end{figure}

\begin{figure}
    \centering
    \includegraphics[]{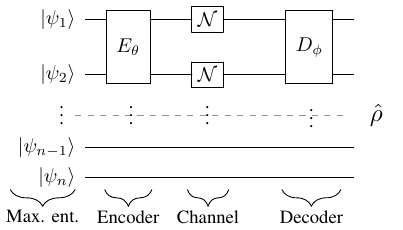}
    \caption{The circuit model for quantum communication. A maximally entangled quantum state is taken as input where part of the system is fed through a parameterized encoder, the noise model, and a parameterized decoder. The density matrix representing the final state is taken as output. The red dashed line represents the system separation where some of the state remains with the sender.}
    \label{fig:qc-coder}
\end{figure}

In information theory, channel capacity theorems are often proven non-constructively, leaving the challenging task of determining a capacity-achieving code open-ended. This holds true for quantum communication as well, where, for example, the Holevo capacity outperforms classical capacity \cite{wilde2013quantum} for certain channels but joint-detection receivers (JDRs) that are easy to produce is still a challenging problem~\cite{banaszek2020quantum, guha2011structured}. By using quantum autoencoders for training communication systems, we believe it can lead to easier-to-build JDRs that can allow for higher communication rates approaching the theoretical limits. This is highly aligned with the objective of future communication systems, especially in 6G networks. 6G networks aim to incorporate quantum features into the communication networks to enhance performance. Using our framework, we can, for example, learn the optimal channel codes under the constraints of what 6G networks can do.

In previous literature, autoencoders have been extensively used in a classical setting to generate channel codes for noisy classical channels~\cite{o2017introduction, letizia2021capacity, jiang2019turbo}. While these concepts have been explored in the classical domain, their application to quantum communication remains unexplored. Hence, in this work, we investigate this unexplored direction. Our approach involves developing a framework to analyze quantum channel codes and using it to evaluate the performance of learned codes across various channels. We focus on three communication settings: 1) Classical-quantum communication; 2) Incorporating shared entanglement resources into the model for entanglement-assisted (EA) communication; and 3) Quantum communication. Our investigations yield promising results, with the models efficiently learning encoding and decoding parameters and approaching theoretical capacities in all cases.

\section{Channel Coding with Quantum Autoencoders}

We explore the performance of parameterized circuits training for learning classical, and EA-classical, and quantum capacities for various channels. Each communication setting has a slightly different configuration but there is a general structure that allows us to easily modify the framework to accommodate. 

In the classical communication setting, the communication process is the following: a sender chooses a classical message from a codebook $s \in \mathcal{M}$, which is then encoded into a quantum state $\rho_s$ using a unitary operation $E_\theta$. The parameterized operation $E_\theta$ has a particular circuit structure and a $k$-length vector $\theta \in \mathbb{R}^k$ defines the operations. The encoded state $\rho_s$ is transmitted through a channel $\mathcal{N}$, a completely positive, trace preserving map. The resulting state $\mathcal{N}(E_\theta(\rho_s))$ is processed by a decoder $D_\phi$, which is parameterized by $\phi \in \mathbb{R}^{k'}$, where $k'$ can be different than $k$. The measurement of the state yields a classical message $\hat{s}$ approximating $s$ and defines a conditional distribution $p(\hat{s}|s)$. This setting is illustrated in Fig.~\ref{fig:coder}. In some cases, we introduce redundant qubits to test repetition codes, for example. In these cases, a parameterized ``pooling'' circuit (Fig.~\ref{fig:pooler}) is incorporated in the decoder, placed between $D_\phi$ and the measurements. To analyze the quality of the learned channel code, we calculate the respective mutual information using input-output values over the channel and compare it to the known channel capacities.  

For EA communication, additional qubits are introduced to the system and are entangled with the communication system before communication occurs. In this case, we assume the entanglement is distributed in a noiseless way, but we do not presume which state the entanglement was in to start with. Pairs of qubits are therefore firstly entangled using parameterized entangling operations with a gate set forming $K_\lambda$, where $\lambda$ is the parameter vector. For the encoding step, the classical message is fed into the encoder such that controlled gates can manipulate the sender's half of the entangled state. The framework applies channel noise to the sender's half and the other half of the system does not experience the channel noise. The receiver then uses the total system for message decoding. This setting is illustrated in Fig.~\ref{fig:ea-coder}. Again to analyze the quality of the learned EA channel code, we calculate the respective mutual information. 

The quantum capacity of a quantum channel is defined in a different way than classical capacities are defined. In this case, it is not a direct measure of which rate quantum states are being transmitted, but rather the direct value can be thought of as a measure of how well a quantum channel preserves entanglement. To therefore compute the quantum capacity of a quantum channel, the communication task is to firstly generate a maximally entangled bi-partite system and then apply the encoding, channel model, and decoding to one half of the system. The value in which to maximize is no longer the mutual information, but rather the quantum coherent information defined as $I(A\rangle B)_{\rho} \coloneqq S(B)_{\rho} - S(AB)_{\rho}$, where $S(X)_{\rho}$ is the von Neumann entropy of state $\rho$ restricted to subsystem $X$. To simulate this in the framework, we apply parameterized encoding and decoding gates to the subsystem that undergoes channel noise as depicted in Fig.~\ref{fig:qc-coder}.

In this initial work, we have set up the framework for learning quantum channel codes and analyzed learned codes for different channels. In particular, we study the quantum bit-flip, phase-flip, depolarizing, and amplitude damping channel for $p=1/2$ and $p=1$. The Kraus operations modeling the behavior for these channels are the following:
\begin{enumerate}
    \item \textbf{Bit-flip}:
    \begin{align}
        K_0 = \sqrt{1-p}\begin{bmatrix}1 & 0 \\ 0 & 1 \\ \end{bmatrix}, K_1= \sqrt{p}\begin{bmatrix} 0 & 1 \\ 1 & 0 \\ \end{bmatrix}
    \end{align}
    \item \textbf{Phase-flip}:
    \begin{align}
        K_0 = \sqrt{1-p}\begin{bmatrix} 1 & 0 \\ 0 & 1 \\ \end{bmatrix}, K_1= \sqrt{p}\begin{bmatrix} 1 & 0 \\ 0 & -1 \\ \end{bmatrix}
    \end{align}
    \item \textbf{Depolarizing}:
    \begin{align}
        K_0 = \sqrt{1-p}\begin{bmatrix} 1 & 0 \\ 0 & 1 \\ \end{bmatrix}, K_1&= \sqrt{p/3}\begin{bmatrix} 1 & 0 \\ 0 & -1 \\ \end{bmatrix} \\\nonumber
        K_2 = \sqrt{p/3}\begin{bmatrix} 0 & 1 \\ 1 & 0 \\ \end{bmatrix}, K_3&= \sqrt{p/3}\begin{bmatrix} 0 & -i \\ i & 0 \\ \end{bmatrix} 
    \end{align}
    \item \textbf{Amplitude Damping}:
    \begin{align}\nonumber
        K_0 = \sqrt{1-p} \begin{bmatrix} 1 & 0 \\ 0 & \sqrt{1-\gamma} \\ \end{bmatrix}, K_1&= \begin{bmatrix} 0 & \sqrt{\gamma(1-p)} \\ 0 & 0 \\ \end{bmatrix} \\
        K_2 = \sqrt{p}\begin{bmatrix} \sqrt{1-\gamma} & 0 \\ 0 & 1 \\ \end{bmatrix}, K_3 &= \begin{bmatrix} 0 & 0 \\ \sqrt{\gamma p} & 0 \\ \end{bmatrix} 
    \end{align}
\end{enumerate}

\begin{figure}
    \centering
    \includegraphics[scale=1]{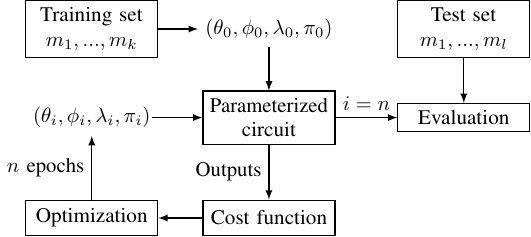}
    \caption{A flow diagram of how the framework executes. For classical message transmission a training collection of messages $\{m_i\}_{i=1}^k$ are selected and the initial parameters $(\theta_0, \phi_0,\lambda_0, \pi_0)$ are set. The data is fed into the parameterized circuit which outputs either classical measurement results or a density matrix. The cost function is computed and the parameters are updated. This repeats $n$ times. After $n$ iterations, the final parameters are used to evaluate the code against a test set of messages, when applicable for classical communication.}
    \label{fig:system-model}
\end{figure}

We configure the model by combining a parameterized encoder, channel, and parameterized decoder, along with optional components such as the pooling layer. This configuration can be trained using the following software tools. Our framework leverages the JAX interface \cite{frostig2018compiling} of the Pennylane software library \cite{bergholm2018pennylane} for quantum simulation. For parameter updating, we employ a variant of the Adam optimizer \cite{DBLP:journals/corr/abs-1908-03265}. In the classical communication case, we employ the average cross-entropy loss \cite{good1952rational} as the cost function. In the quantum communication case, we use the trace distance between the output state of the reference system. The training flow is depicted in Fig.~\ref{fig:system-model}. Once the model is trained for a specific channel, a set of test messages is used to evaluate the code. In the classical case, we compute the mutual information, while in the quantum case, we calculate the quantum coherent information. 

\section{Training Results}


In this section, we present the results of training the circuit models and the evaluation of the final trained codes in their ability to achieve capacity. We explore the classical and quantum communication scenarios.

\subsection{Classical Communication}\label{sec:cc}

Classical communication over quantum channels can be done in a variety of ways and therefore there are four main capacities to consider when no additional entanglement resources are used: 1) Separable-state encoding and decoding $C_{ss}$; 2) Separable-state encoding and joint measurement decoding $C_{sj}$, also known as the Holveo capacity \cite{holevo1998capacity}; 3) Entangled-state encoding and separable measurement decoding $C_{es}$; and 4) Entangled-state encoding and joint decoding $C_{ej}$ \cite{liang2002entanglement}. Depending on the desired model, our framework allows to test any of the cases. 

For the demonstration of our framework in the classical communication setting without entanglement, we have tested various code models for the (a) Bit-flip channel, (b) Depolarizing channel, and (c) $p=1$ amplitude damping channel. The results of training the code are shown in Fig.~\ref{fig:caps}(a)-(c). For the bit-flip and depolarizing channel, the framework was able to train the circuit parameters to produce a code that meets the channel capacity. We can see that by adding an encoding operation to the bit-flip channel, the Holveo capacity is exceeded since it becomes possible to encode in the $X$ basis. For the depolarizing channel, encoding at the sender's side does not enhance the capacity of the channel in this case and the Holveo capacity is equal to $C_{ej}$. 

The $p=1$ amplitude damping channel is a non-unital channel, which means that the accessible information between channel uses is non-additive, making the determination of channel capacity more challenging. For the $p=1$ amplitude damping channel, the Holveo quantity serves as a lower bound for the capacity. Initially, without encoding on the sender side and using a standard basis embedding, the output of our framework produced a code with a sub-optimal capacity, lower than the lower bound in the single qubit case. To further explore this, we introduced repetition and pooling techniques (see Fig.~\ref{fig:pooler}) in the encoding and decoding scheme. Although the repetition code uses the channel more times, the circuit with pooling yields a single output bit. We compute the mutual information (without regularization) using this single bit and observe that for $\gamma$ values greater than 0.5, repetition and pooling improve the results, where the solid lines in the plot represent the no-encoding scheme. When we introduce a parameterized encoding circuit (represented by the dashed lines) with the same pooling scheme, we observe a significant improvement in the mutual information of the code. We present known upper and lower bounds for the total capacity of the channel using the dash-dotted lines.

\subsection{Entanglement-Assisted Classical Communication}\label{sec:ea}

\begin{figure*}
    \centering
    \includegraphics[]{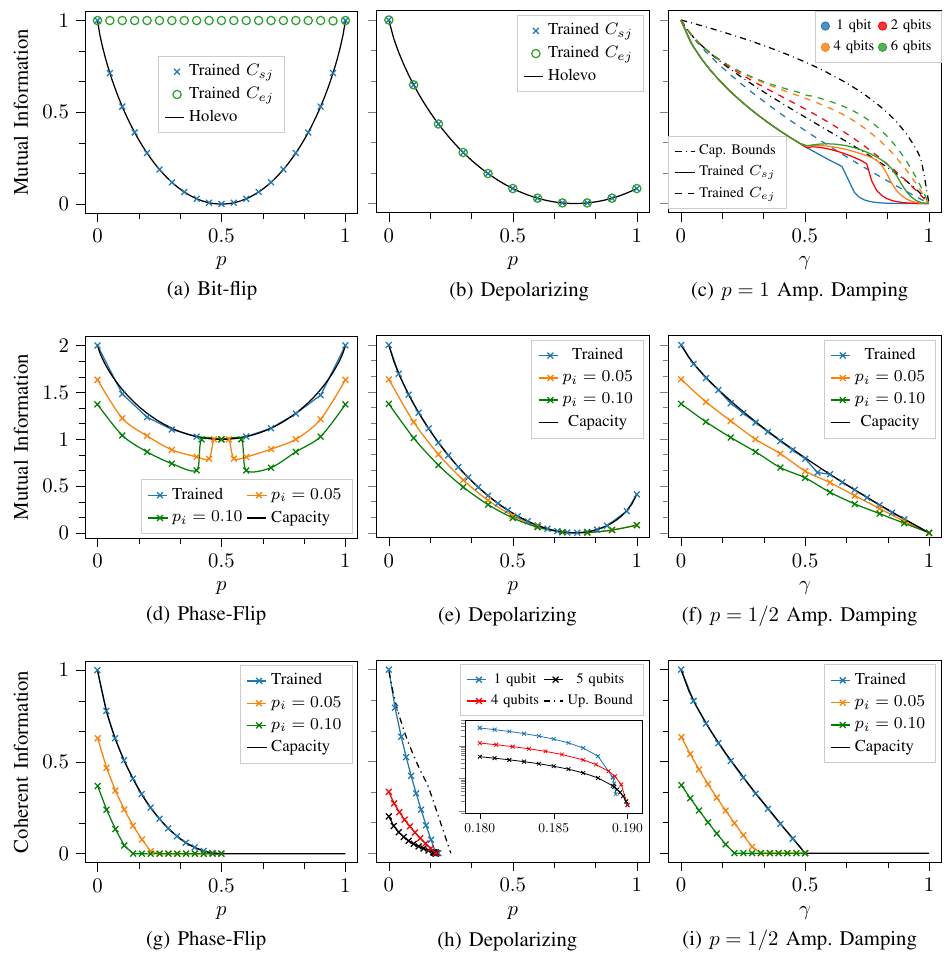}
    \caption{The training results for various channel codes for classical, EA-classical, and quantum communication produced by the framework for various channel models. In plots (a) and (b), we show with green circles how adding an encoding layer affects the code capacity. In plot (c), we show the results of using a repetition code with a pooling layer. In the plots with orange and green curves from (d)-(i), we show how memory decoherence on the sender's side affects the code capacity using depolarizing noise with idler probability $p_i$. In (h), we show how GHZ states of varying sizes displays super-additivity in the zoomed-in plot.}
    \label{fig:caps}
\end{figure*}

When considering entanglement-assisted (EA) communication, important considerations include determining the type of entanglement to be used (e.g., Bell states, GHZ states, etc.) and how these states should be encoded with classical information. In our framework, we delegate these complex decisions to the training process.  The model in this case, as depicted in Fig.~\ref{fig:ea-coder}, parameterizes the initial entanglement generation and then the classical encoding step. Then, part of the system gets the noise effects from the channel and some remains noiseless. Upon arrival, the receiver decodes the total state also using a parameterized circuit. We analyzed the EA capacity for three channels: (a) Phase-flip, (b) Depolarizing, and (c) $p=1/2$ amplitude damping. The results in Fig.~\ref{fig:caps}(d)-(f) show that the framework is able to train the parameterized circuit model to find the optimal parameters that nearly meet or do meet the channel capacity in each case. In these cases, the channel models are unital channels and so their capacities are known. They are found by using standard super-dense coding which our trained model successfully reproduced.  

In these results, we assumed that the stored entanglement remains noiseless while the sender makes transmissions, but we also test the case where a small amount of depolarizing noise affects the stored qubit. 
In Fig.~\ref{fig:caps}(d)-(f), represented by the orange and green crosses, we also show the result of depolarizing noise applied to the idle qubit to simulate the effects of memory. We assume the sender's qubit undergoes a depolarizing channel with $p_i=0.05$ and $p_i=0.10$ during transmission. We see that the noisy memory diminishes the rate and that the framework could still train the model to achieve the maximum rate. We show the results in Fig.~\ref{fig:caps}(d)-(f) with red crosses.

\subsection{Quantum Communication}\label{sec:qc}

The final communication scenario we considered for our framework is the quantum communication setting. In this scenario, we focused on transmitting parts of maximally entangled states and evaluating the preservation of entanglement. Based on the configuration in Fig.~\ref{fig:qc-coder}, we set the model for training. In quantum communication, channels that are ``degradable'' have a single-letter quantum capacity formula~\cite{devetak2005capacity}, making them easier to analyze. The phase-flip and $p=1/2$ amplitude damping channels are such channels, where the depolarizing channel is not. We show the training results for determining the quantum capacity for those channels in \ref{fig:caps}(g)-(i). Notably, for degradable channels, a single maximally entangled state meets the capacity, and our framework successfully reproduces this result.  However, the capacity for the depolarizing channel is generally unknown. We explore various methods of transmitting entanglement over this channel. The blue crosses in Fig.~\ref{fig:caps}(h) show the single unit of entanglement. The red and black crosses show the regularized performance of transmitting $n-1$ parts of an $n$-qubit GHZ state. As observed, the performance is worse, but intriguingly, for larger values of $p$, this form of entanglement outperforms the single-qubit case, which is already a known fact. To highlight this advantage, we provide a zoomed-in plot within (b) on a logarithmic scale, revealing that the 4- and 5-qubit cases remain positive for larger values of $p$. As in the entanglement-assisted case, we show the effects of memory noise for at the sender in (g) and (i) on the capacity.

In summary, there is much more to learn about quantum capacities, and our framework serves as a valuable tool for deeper investigations. The distribution of entanglement is a fundamental aspect of future quantum networks, and understanding the ultimate rates of entanglement distribution holds significant importance.

\section{Conclusion and Outlook}

In conclusion, we have developed a quantum machine learning framework for training quantum channel codes over different channel models and constraints. Our framework has demonstrated its effectiveness in learning capacity-achieving codes for qubit channels. We have modeled classical, entanglement-assisted classical, and quantum capacities, and in each case, our framework has shown strong performance. The trained models have achieved capacities that are close to the known limits for unital channels, enabling straightforward analysis of various error-correcting codes. Furthermore, our framework has allowed us to observe quantum effects, such as super-additivity, with just a few lines of software code.

Although we have covered a vast subset of theory in this work, we strongly believe that quantum machine learning holds tremendous potential for future quantum communication systems, particularly in the era of 6G networks where strong constraints will have to be imposed to include quantum features. Our framework can be extended in numerous ways to explore various communication scenarios and channel capacities, including private and zero-error capacities, as well as adaptability to multiple input-output channels. Additionally, we foresee promising outcomes by exploring different pooling strategies~\cite{monnet2023pooling} to facilitate efficient joint-detection receiver designs. Furthermore, while this work focuses on qubits, conducting an in-depth study of continuous variable communication can yield novel insights for optical communication, and is a project we are currently undertaking. In conclusion, our research introduces a new domain of investigation, employing ML techniques for quantum communication, and establishes the foundation for future explorations.

\section*{Acknowledgements}
The authors thank Bing Qi, Hassan Shapourian, and Ionel Miu for helpful discussions.

\clearpage
\clearpage

\bibliographystyle{IEEEtran}
\small{\input{refs.bbl}}

\end{document}